\documentclass[12pt,preprint]{aastex}

\shorttitle{Debris around Vega}
\shortauthors{Wilner et al.}

\begin{document}

\slugcomment{to appear in ApJ Letters}

\title{Structure in the Dusty Debris around Vega\footnote{Based on 
observations carried out with the IRAM Plateau de Bure Interferometer. 
IRAM is supported by INSU/CNRS (France), MPG (Germany) and IGN (Spain).}
}

\author{D.J.\ Wilner, M.J. Holman, M.J. Kuchner\altaffilmark{2}, and P.T.P. Ho}

\email{dwilner, mholman, mkuchner, pho@cfa.harvard.edu}

\affil{Harvard-Smithsonian Center for Astrophysics, 
60 Garden Street, Cambridge, MA 02138}

\altaffiltext{2}{Michelson Postdoctoral Fellow}

\begin{abstract}
We present images of the Vega system obtained with the 
IRAM Plateau de Bure interferometer at 1.3 millimeters wavelength
with sub-mJy sensitivity and $\sim2\farcs5$ resolution (about 20 AU).
These observations clearly detect the stellar photosphere 
and two dust emission peaks offset from the star by $9\farcs5$ and
$8\farcs0$ to the northeast and southwest, respectively.
These offset emission peaks are consistent with the barely resolved 
structure visible in previous submillimeter images, and 
they account for a large fraction of the dust emission.
The presence of two dust concentrations at the observed 
locations is plausibly explained by the dynamical influence of 
an unseen planet of a few Jupiter masses in a highly eccentric orbit
that traps dust in principal mean motion resonances.    

\end{abstract}


\section{Introduction}
The Infrared Astronomical Satellite (IRAS) measured a 
far-infrared flux from the vicinity of the main sequence star Vega
greatly in excess of the photospheric emission (Aumann et al.\ 1984).
We now know that main sequence
stars with excess infrared emission, the ``Vega-excess'' stars,
are common, and that this phenomenon represents
circumstellar dust particles heated by stellar radiation
(see the reviews by Backman \& Paresce 1993, 
Lagrange, Backman \& Artymowicz 2000, and Zuckerman 2001).
Because dust destruction by Poynting-Robertson (P-R) drag 
and sublimation happens on a timescale shorter than 
main-sequence lifetimes, the dust orbiting Vega-excess stars must be 
continually replenished by the collisions of larger orbiting bodies, 
possibly analogous to Kuiper Belt objects. 

Imaging and photometry of Vega-excess stars suggest that these systems 
commonly contain clouds of dust and debris shaped like disks or rings 
(e.g. Smith \& Terrile 1984, Jayawardhana et al.\ 1998, Koerner et al.\ 1998).
Some images contain hints that massive planets may help to sculpt the clouds;
many are evacuated interior to a radius of 10 to 80 AU.  
Recent observations at 850~$\mu$m with the James Clerk Maxwell Telescope 
(JCMT) have provided the first detailed images of the dust emission 
around several of these stars, including Vega (Holland et al.\ 1998, 
Greaves et al.\ 1999, Dent et al.\ 2000).   
At $14''$ resolution, the 850~$\mu$m images of Vega, $\epsilon$~Eridani, 
Fomalhaut, and $\beta$~Pictoris reveal emission peaks offset from the stars.  
The origin of these peaks is not clear. Some authors have modeled 
them as concentrations of dust trapped by planets, assuming planets 
on roughly circular orbits (Liou \& Zook 1999, Ozernoy et al.\ 2000)

Millimeter wavelength interferometry offers a way to obtain high-resolution
information on the emission structures around nearby Vega-excess stars,
including the central cavities and mysterious offset emission peaks.
Vega, an A0V star just 7.76~pc away, is positioned favorably in the 
northern sky for existing millimeter wavelength interferometers, 
and it appears to be viewed nearly pole-on,
which simplifies the interpretation of its images (Gulliver et al.\ 1994).
Recently, Koerner, Sargent \& Ostroff (2001) presented an image of Vega 
at 1.3~mm from the Owens Valley Radio Observatory (OVRO) that
resolves several emission peaks confined to a circumstellar ring.
In this {\em Letter}, we present new interferometric observations of Vega 
with better sensitivity that reveal intriguing asymmetries in the 
locations of the emission peaks.
These asymmetries may be the dynamical signature of a planet 
of a few Jupiter masses in a highly eccentric orbit.

\section{Observations}
We observed Vega in the 1.3~mm and 3~mm bands simultaneously 
with the IRAM Plateau de Bure interferometer (PdBI).
Table 1 summarizes the observational parameters. 
The total integration time on source was about 23 hours, 
in excellent weather with precipitable water vapor content less 
than 3~mm and r.m.s.~phase errors less than $30^{\circ}$ at 1.3~mm. 
Flux densities were set with reference to the standard source MWC349,
and the systematic uncertainties in the flux scale are estimated 
to be about 20\%. 

\section{Results}

\subsection{1.3~mm continuum}
\begin{figure}
\epsscale{1.0}
\plotone{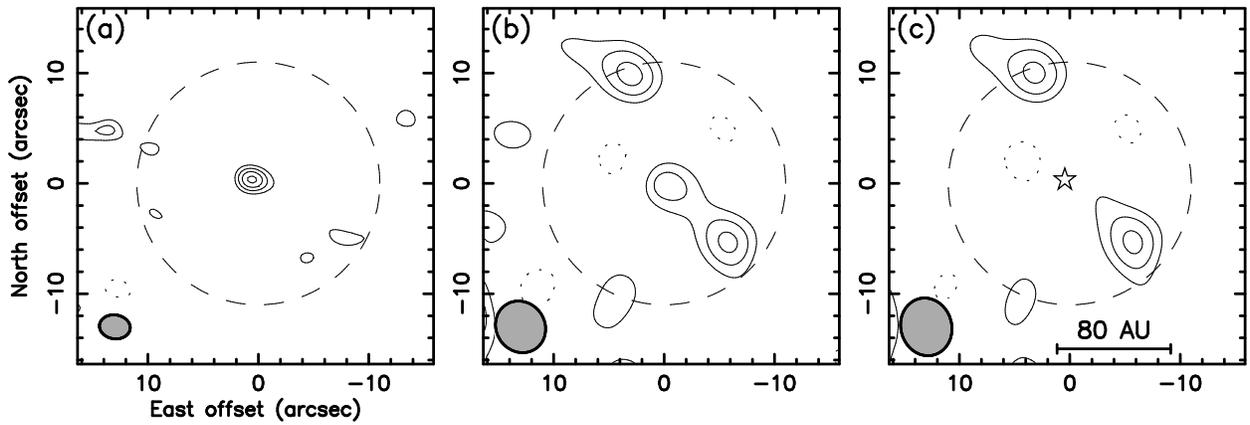}
\caption{
Three 1.3~mm images of the Vega system 
(not corrected for the primary beam response). 
(a) High resolution image showing the stellar photosphere.
The shaded ellipse shows the 
$2\farcs8\times2\farcs1$ p.a.~$80^{\circ}$ synthesized beam. 
The contour levels are $\pm2,3,4,5\times0.30$~mJy.
Negative contours are dotted.
The dashed circle denotes the $22''$ primary beam half power field of view.
(b) Low resolution image to emphasize extended emission features.
The shaded ellipse shows the 
$4\farcs8\times4\farcs4$ p.a.~$35^{\circ}$ synthesized beam. 
The contour levels are $\pm2,3,4\times0.52$~mJy.
(c) Lower resolution image with the stellar contribution subtracted.
The star symbol markes the position of the star.
The shaded ellipse shows the 
$5\farcs3\times4\farcs6$ p.a.~$15^{\circ}$ synthesized beam. 
The contour levels are $\pm2,3,4\times0.57$~mJy.
}
\label{fig:vega-images}
\end{figure}

Figure~\ref{fig:vega-images} shows three 1.3~mm images of 
the Vega field that emphasize features in the visibility data 
of different spatial extent and surface brightness. 

Figure~\ref{fig:vega-images}a shows a high resolution view
($2\farcs8\times2\farcs1$),
which is dominated by emission from the stellar photosphere
at a position consistent with the image center 
(see Downes et al.\ 1999 for a discussion of astrometric errors).
A least squares point source fit to the visibility data 
gives flux density $1.7\pm0.13$ mJy.
Given the systematic uncertainties in the flux scale, 
the measured flux density is consistent with the 2.3~mJy 
expected from an extrapolation to longer wavelengths 
of the photospheric model of Cohen et al. (1992). 
The uncertainties also accommodate a small contribution 
from warm dust within 5~AU of the star, suggested by 
near-infrared interferometry (Ciardi et al.\ 2001).
The detection of the unresolved star provides an excellent 
internal check on the calibration and data quality.

Figure~\ref{fig:vega-images}b shows a lower resolution image 
obtained by applying a taper to the visibility data 
to increase the surface brightness sensitivity
($4\farcs8\times4\farcs4$).
Two additional emission peaks become apparent, 
one offset $8\farcs0$ to the southwest of the star 
(position angle $224^{\circ}$), 
and the other offset $9\farcs5$ to the northeast of 
the star (position angle $13^{\circ}$).
The uncertainties in the peak locations and position angles 
are $\pm0\farcs5$ and $\pm10^{\circ}$, respectively.
The northeast-southwest orientation of the peaks corresponds 
closely to the orientation of the barely resolved structure 
in the JCMT image at 850~$\mu$m (Holland et al.\ 1998),
and the positions of the peaks are similar to the main significant 
features in the OVRO image at 1.3~mm (Koerner et al.\ 2001).
The OVRO image shows two additional peaks along an arc 
northeast of the star that are not apparent in the PdBI data.
The origin of this difference is not clear, but it may reflect the
inherent difficulties in low signal-to-noise imaging of extended 
emission near the edge of the field of view. More data with better 
sensitivity are needed to verify the reality of these extra features.

Figure~\ref{fig:vega-images}c shows an image made with a
more extreme taper ($5\farcs3\times4\farcs6$ p.a. $15^{\circ}$),
after subtracting from the visibility data the point source 
contribution of the central star.
This image best isolates the two offset emission peaks.
The fluxes obtained from fitting two elliptical Gaussians 
to the visibilities are $7.1\pm1.4$~mJy and $4.3\pm1.0$~mJy 
for the northeast peak and southwest peak, respectively. 
The quoted uncertainties include the formal errors in 
the fits and corrections for primary beam attenuation, 
but not the systematic errors in the flux scale.
The summed flux in the peaks of $11.4\pm1.7$~mJy is
consistent with that measured from the OVRO image, 
and it represents a sizeable fraction of the total dust emission
in the system by extrapolation from the 850~$\mu$m data.
Assuming standard dust mass opacities (see Holland et al.\ 1998), 
the mass of emitting dust in each peak is less than 0.3 lunar masses. 

\subsection{3.3~mm continuum}
No significant emission at 3.3~mm is detected from either 
the star or the offset peaks, with an r.m.s.~noise level 
of 0.13~mJy in a $7\farcs8\times4\farcs8$ beam. 
These upper limits are consistent with 
a $\nu^2$ blackbody spectrum for the star and
a steeper spectrum for the dust peaks.

\subsection{CO J=2--1 line}
The channel maps shows no evidence for CO J=2--1 line emission. 
The limit is 0.05 K ($3\sigma$) in a 12~km~s$^{-1}$ 
velocity bin and $2\farcs8\times2\farcs1$ beam. 
This limit rules out beam dilution as an explanation for previous 
nondetections of CO emission (Dent et al.\ 1995).
As discussed by Kamp \& Bertoldi (2000), the CO molecules in 
the Vega environment are likely destroyed by photodissociation, 
and the CO emission does not reliably trace molecular gas content.

\section{Discussion}
The new high resolution 1.3~mm imaging is consistent with 
the inference from previous observations that the Vega debris 
takes the form of a clumpy ring viewed nearly pole-on.  

Several explanations for the emission peaks are viable.
They could conceivably represent dust clouds released by 
recent collisions of planetesimals. The collisions would have to be 
recent since such clouds would disperse in $\sim$10--100 orbital periods,
and the planetesimals must be massive enough that their collisions 
release a substantial fraction of a lunar mass in dust.
It has been suggested that the emission peaks may represent 
planets which appear larger than they are because they have 
somehow retained circumplanetary disks (Holland et al.\ 1998).  
The presence of two emission peaks in the Vega system would call 
for two such planets (or one planet and one cloud).
Alternatively, the peaks could be due to unrelated background galaxies, 
though this is very unlikely given recent source counts 
(Blain et al.\ 1999, Chapman et al.\ 2002).

A more likely scenario is that the two emission peaks represent 
dust clumps created by the dynamical influence of an unseen planet
or planets.
For example, in our Solar System, the Earth (and probably Neptune)
temporarily traps interplanetary dust as it spirals inward via 
P-R drag in a series of first-order mean motion
resonances (MMRs) to produce a ring of enhanced dust density along the 
Earth's orbit (Dermott et al.\ 1994, Liou \& Zook 1999).
Astrometric limits exclude a companion to Vega more massive 
than 12 Jupiter masses with a period less than 7 years 
(Gatewood \& de Jonge 1995), and imaging searches near Vega have 
failed to detect objects as bright as 12 Jupiter mass brown dwarfs
(Holland et al.\ 1998).  However, a less massive companion-- 
still massive enough to thoroughly reorganize a debris cloud--
may yet have escaped detection.

Most of the known extrasolar planets do not resemble Earth or Neptune; 
they are more massive than Saturn and often have significant 
orbital eccentricities (Marcy \& Butler 2000).  
Models of Jupiter mass planets on eccentric orbits interacting with 
inward-spiraling dust particles suggest that such a planet often does not 
create a ring, but may create a pair of orbiting dust clumps 
(Kuchner \& Holman 2001).  
The two dust enhancements are generally not co-linear with the star, 
and typically one is farther from the star than the other.  
We suggest that if the emission peaks in the PdBI images of Vega represent 
dust concentrations created by the dynamical influence of a planet, then 
their asymmetries point to a Jupiter mass planet with an eccentric orbit. 

Low mass planets and planets with nearly circular orbits trap dust 
in resonances near the planet (Roques et al.\ 1994, Ozernoy et al.\ 2000).  
For a Jupiter mass planet with an eccentric orbit, more distant
resonances may trap significant quantities of dust.  
The principal resonances, where the planet orbits roughly $n$ times for 
every one orbit of the particle, are the strongest of the distant MMRs. 
The 2:1, 3:1, 4:1, and 5:1
resonances are located at semimajor axes $a\approx$1.59, 2.08, 2.52,
2.92 times the semimajor axis of the planet's orbit, respectively.

Dust approaching a massive, eccentric planet is typically trapped in 
the exterior principal resonances with resonant arguments of the form:
\begin{equation}
\sigma = (k+1)\lambda - \lambda_p - (k-1)\varpi - \varpi_p,
\end{equation}
where $k=1, 2, 3,...$, $\lambda$ and $\lambda_p$ are the mean
longitudes, and $\varpi$ and $\varpi_p$ are the longitudes of
periastron of the particle and planet, respectively.  
Take the longitude of periastron of the planet as the reference
angle ($\varpi_p\equiv 0$).  In these resonances, 
$\sigma$ oscillates about zero.  These terms share a peculiar property.   
When a particle is at periastron ($\lambda = \varpi$), 
$\sigma = 2\varpi - \lambda_p \sim 0$.  So, when the planet
has mean longitude $\lambda_{p}$, trapped particles with longitudes 
of periastron at $\varpi=\lambda_{p}/2$ and $\varpi=\lambda_{p}/2+\pi$
must be near periastron.  This resonance condition creates 
two dust concentrations which appear to revolve around the star
at half the planet's orbital frequency.  
The patterns from different resonances (e.g. 3:1 and 4:1) occur 
at the same longitude and reinforce each other.  
The concentrations are only wave-like patterns in the dust distribution;  
the actual particles orbit more slowly than the pattern.
Secular and resonant effects on the particles' eccentricities
and longitudes of periastron create the asymmetries between 
the two concentrations of dust.

To illustrate how a cloud of particles in these principal resonances 
might appear, we created a model image by numerically integrating 
the orbits of 500 test particles under the influence of gravity from Vega 
and a single planet, radiation pressure, and P-R~drag.  We performed the 
integration with a symplectic $n$-body map (Wisdom \& Holman 1991) to which 
we added terms representing radiation pressure and P-R~drag, 
a dissipative force (Cordeiro et al.\ 1996, Mikkola 1998, Kehoe 2000).

The effect of stellar radiation on dust particles is parameterized by 
$\beta$, the ratio of the radiation pressure force to the gravitational force 
(Burns et al.\ 1979).
For spherical particles with density 2 g~cm$^{-3}$ orbiting Vega 
($M_0=2.5~M_{\odot}$, $L_0=60~L_{\odot}$), $\beta = 6.84/s$, 
where $s$ is the particle radius in $\mu$m.   
The IRAS spectral energy distribution of Vega suggests that particles 
smaller than $\sim 80~\mu$m contribute little to the emission.
For this model, we chose $\beta = 0.01$, which 
corresponds to a particle size of $\sim685~\mu$m.

The planet was given a mass of 3 Jupiter masses, an orbit with
a semimajor axis of 40~AU, and an orbital eccentricity of 0.6.  
The test particles were placed near 108~AU from Vega with small free
eccentricity ($e\sim 0.2$) and small free longitude of periastron 
($\varpi\sim\varpi_p$). The particles were
also given slight initial inclinations ($ i\sim 4^\circ$) and random
longitudes of ascending node.  The integrations ran for $4\times 10^8$
years, the estimated age of Vega, using a step size of 5 years.  
A typical millimeter sized particle was ejected by the planet in 
$\sim 10^8$~years after capture into resonance. 

The role of collisions among dust particles, ignored in our models, 
is uncertain.  The optical depth of brightest dust concentration is 
$\sim 10^{-3}$, which implies a collisional timescale of $\sim 10^7$ years.
This is shorter than the P-R timescale of $\sim 10^8$ years for millimeter 
sized grains, so such grains could not migrate far from their source 
before a collision. However, this optical depth is not representative of the 
whole disk. Furthermore, for a steady-state size distribution that results 
from a collisional cascade, most of the mass is concentrated in the larger 
bodies while most of the area is concentrated in the smaller bodies 
(Dohnanyi 1969). Most collisions are not catastrophic. Grains that are 
broken up are fragmented by a particles with just enough mass to do so.  
Thus the collisional fragments have velocities nearly identical to the 
parent body (Wyatt et al.\ 1999).  Therefore, the collisional timescale
above is likely an underestimate.  Nevertheless, if the collisional timescale 
is short, then this argues for source debris which is itself near or in a 
principal resonance. A resonant population of source bodies could be 
trapped by a migrating planet, as Neptune may have trapped the plutinos 
(Malhotra 1995).  Here we consider only trapping via the migration of 
dust particles, in order to illustrate the basic geometry of the resonant 
structures.

\begin{figure}
\epsscale{1.0}
\plotone{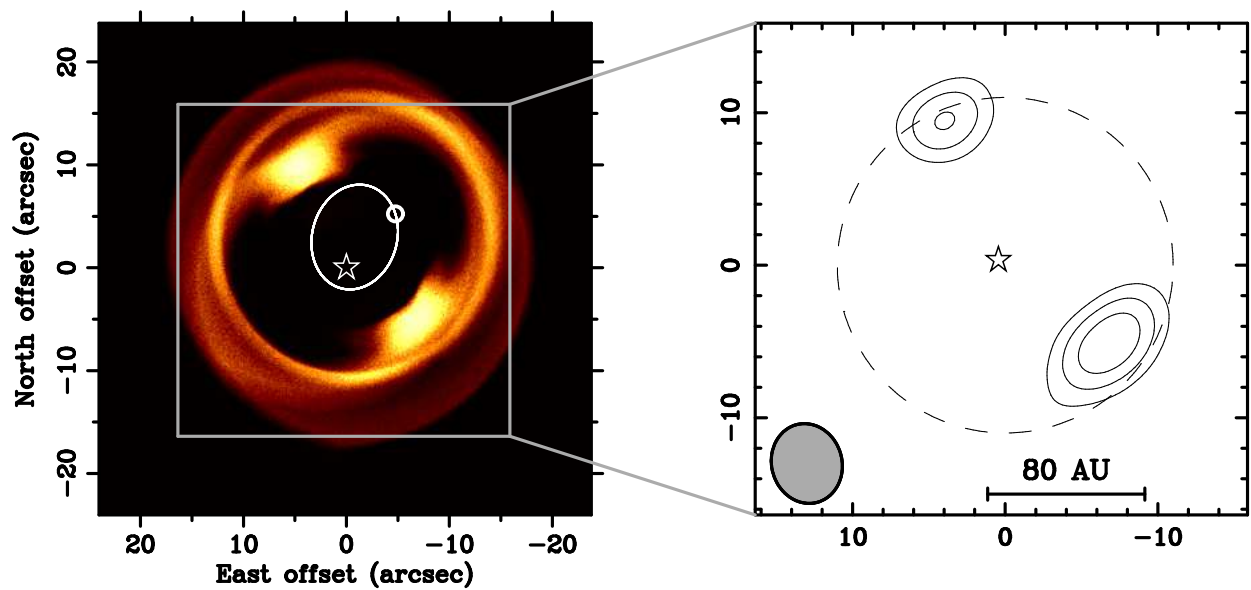}
\caption{
{\em (left)} 
A representative numerical simulation of 1.3~mm dust emission from 
orbital dynamics that includes a 3~Jupiter-mass planet, radiation pressure, 
and Poynting-Robertson drag. The dust becomes temporarily entrained 
in mean motion resonances associated with the planet,
producing a prominent two-lobed structure.
The ellipse represents the planet's orbit, and the circle 
marks the position of the planet.
{\em (right)} Simulated observation of the numerical model, 
taking account the PdBI response for the Vega observations.
The percentage contour levels are the same as in Figure~\ref{fig:vega-images}c.
}
\label{fig:vega-model}
\end{figure}

\begin{figure}
\epsscale{0.5}
\plotone{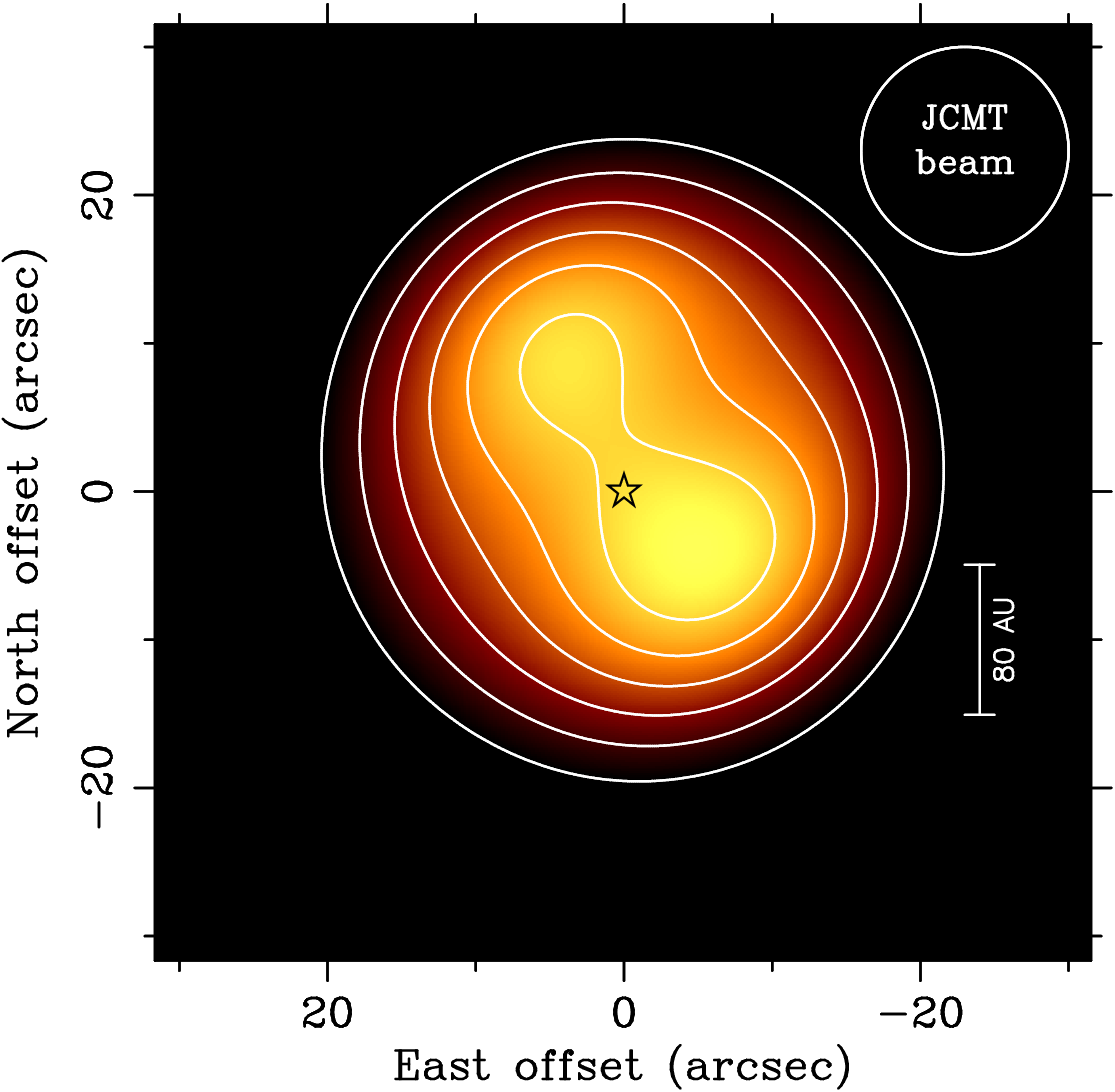}
\caption{
{\em (left)} 
The model image from Figure~\ref{fig:vega-model} convolved 
to match the resolution of the JCMT image of Holland et al. (1998)
after scaling the emission to 850~$\mu$m and adding 
a point source at the center to represent the stellar photosphere.
The contours are at 1.9 mJy intervals starting from 3.8 mJy.
The circle in the upper right shows the JCMT beam size. 
Note that extended emission filtered out by 
the interferometer is visible in this image.
}
\label{fig:vega-model-scuba}
\end{figure}

We simulated snapshots of the dust cloud by creating histograms of the 
particle positions calculated for particular orbital phases of the planet,
and converted these histograms, which model the dust column density, 
into simulated 1.3~mm emission images by multiplying them 
by the Planck emissivity for blackbody particles at the appropriate 
distance from Vega.
The left panel of Figure~\ref{fig:vega-model} shows the simulated 
1.3~mm emission where the planet is at mean anomaly $\sim 100^\circ$.  
At this time, the planet is located 
36.9~AU ($4\farcs8$) west and 40.6~AU ($5\farcs2$) north of Vega.
The right panel of Figure~\ref{fig:vega-model}
shows the result of imaging the model brightness distribution 
using the visibility sampling of the PdBI observations of Vega. 
The spatial filtering property of the interferometer tends to suppress 
the smooth components of the model brightness and to emphasize the peaks.  
The simulated image qualitatively reproduces the observed asymmetries. 
Figure~\ref{fig:vega-model-scuba} shows the same model, 
scaled to 850~$\mu$m using the observed spectral index, 
adding a 5~mJy point source to account for the star, convolved 
to match the resolution of the JCMT image of Holland et al. (1998).
Like the JCMT image, this low resolution view of the model shows 
a central brightening extending in the northeast-southwest direction 
within a nearly circular boundary. 

This numerical model is meant to be representative.  A range of planet 
and dust parameters can capture the main features of the observations.  
In particular, the two dust concentrations would stand out more from 
the extended disk if the parent bodies that generate the dust were 
constrained to lie in the principal resonances.  We defer a detailed 
description of the dynamics and an investigation of the planet 
parameters to future papers. If the planet mass is 3 Jupiter masses, 
then it may be as bright as 18th magnitude in H band (Burrows et al.\ 1997)
and potentially accessible to direct imaging.

Planets found in radial velocity surveys at $a < 3$ AU often follow 
eccentric orbits, but the asymmetric dust ring of Vega may be the 
first sign of a highly eccentric planet at $a > 30$ AU.
Thommes et al.\ (1999) have suggested that Neptune was scattered into 
a highly eccentric orbit early in the life of the Solar System,
and subsequently, its eccentricity was damped by interactions with
the primordial Kuiper Belt.  Perhaps we are witnessing a similar
phase in the evolution of the Vega system.

\acknowledgements
We acknowledge the IRAM staff from the Plateau de Bure and from Grenoble
for carrying out the observations and for their help during the data reduction.
We are especially grateful to Roberto Neri for his assistance.   
Partial support for this work was provided by NASA Origins of Solar Systems
Program Grant NAG5-8195.

\clearpage

\begin{deluxetable}{lll}
\tablecolumns{3}
\tablewidth{0pt}
\tablecaption{Instrumental Parameters}
\tablehead{
\colhead{} & \colhead{3~mm} & \colhead{1~mm} }
\startdata
Observations: & \multicolumn{2}{c}{2001 February 14,18, March 18,27} \\
~      & \multicolumn{2}{c}{D configuration (5 antennas)} \\
Min/Max baseline: & \multicolumn{2}{c}{15 to 80~meters} \\
Pointing center (J2000):  & \multicolumn{2}{c}
                    {$\alpha=18^{h}36^{m}56\fs33$,
                     $\delta=38^{\circ}47''01\farcs3$}\\
Phase calibrators: & \multicolumn{2}{c}{J1848+323, J1829+487} \\
Bandpass calibrator: & 3C273 & 3C273 \\
Flux calibrator: & MWC349 & MWC349 \\
~~~adopted flux: & 1.03 Jy & 1.70 Jy\\
Primary beam HPBW: & $50''$ & $22''$ \\
Synthesized beam HPBW: 
   & $7\farcs8\times4\farcs8$ P.A. $80^{\circ} $
   & $2\farcs8\times2\farcs1$ P.A. $80^{\circ}$ \\
r.m.s. (continuum image): & 0.13 mJy/beam & 0.30 mJy/beam \\
Spectral Line Correlator: & \nodata &  256 channels, 80 MHz \\
~~~species/transition: & \nodata & CO J$=2-1$ \\
~~~frequency:          & \nodata & 230.5380 GHz \\
~~~center LSR velocity: & \nodata & 0 km~s$^{-1}$ \\
~~~channel spacing:  & \nodata & 0.41 km~s$^{-1}$ \\
r.m.s. (line images): & \nodata & 13 mJy/beam \\
\enddata
\label{tab:obs}
\end{deluxetable}

\clearpage

\clearpage

\end{document}